\documentclass[conference]{IEEEtran}
\IEEEoverridecommandlockouts
\usepackage{cite}
\usepackage{amsmath,amssymb,amsfonts}
\usepackage{algorithmic}
\usepackage{graphicx}
\usepackage{textcomp}
\usepackage[table,xcdraw]{xcolor}
\usepackage{booktabs,tabularx}
\usepackage{hyperref}

\def\BibTeX{{\rm B\kern-.05em{\sc i\kern-.025em b}\kern-.08em
    T\kern-.1667em\lower.7ex\hbox{E}\kern-.125emX}}
\begin{document}

\title{Generating Lead Sheets with Affect: A Novel Conditional seq2seq Framework\\

\thanks{This work is funded by Singapore Ministry of Education Grant no. MOE2018-T2-2-161 and SRG ISTD 2017 129, as well as the RIE2020 Advanced Manufacturing and Engineering (AME) Programmatic Fund (No.A20G8b0102), Singapore.}}

\author{\IEEEauthorblockN{Dimos Makris}
\IEEEauthorblockA{\textit{Information Systems,}\\
\textit{Technology, and Design}\\
\textit{Singapore University}\\
\textit{of Technology and Design} \\
Singapore \\
dimosthenis\_makris@sutd.edu.sg}
\and
\IEEEauthorblockN{Kat R. Agres}
\IEEEauthorblockA{\textit{Yong Siew Toh} \\
\textit{Conservatory of Music} \\ 
\textit{National University} \\
\textit{of Singapore}\\
Singapore \\
katagres@nus.edu.sg}
\and
\IEEEauthorblockN{Dorien Herremans}
\IEEEauthorblockA{\textit{Information Systems,}\\
\textit{Technology, and Design}\\
\textit{Singapore University}\\
\textit{of Technology and Design} \\
Singapore \\
dorien\_herremans@sutd.edu.sg}

}

\maketitle

\begin{abstract}
The field of automatic music composition has seen great progress in the last few years, much of which can be attributed to advances in deep neural networks. There are numerous studies that present different strategies for generating sheet music from scratch. The inclusion of high-level musical characteristics (e.g., perceived emotional qualities), however, as conditions for controlling the generation output remains a challenge. In this paper, we present a novel approach for calculating the valence (the positivity or negativity of the perceived emotion) of a chord progression within a lead sheet, using pre-defined mood tags proposed by music experts. Based on this approach, we propose a novel strategy for conditional lead sheet generation that allows us to steer the music generation in terms of valence, phrasing, and time signature. Our approach is similar to a Neural Machine Translation (NMT) problem, as we include high-level conditions in the encoder part of the \textit{sequence-to-sequence} architectures used (i.e., long-short term memory networks, and a Transformer network). We conducted experiments to thoroughly analyze these two architectures. The results show that the proposed strategy is able to generate lead sheets in a controllable manner, resulting in distributions of musical attributes similar to those of the training dataset. We also verified through a subjective listening test that our approach is effective in controlling the valence of a generated chord progression.
\end{abstract}

\begin{IEEEkeywords}
Lead Sheet Generation, Emotion, Valence, seq2seq, Transformer\\
\end{IEEEkeywords}

\section{Introduction}\label{sec1}

Developing computational music generation systems has been the focus of research for many years \cite{hiller1957musical,herremans2017functional,herremans2017morpheus,deliege2006musical}. With the rapid development of deep generative models, their generation results have become hard to distinguish from real-world data in various applications. In the symbolic music generation domain, diverse strategies have been used for a variety of tasks \cite{briot2020deep}. Examples of music generation tasks are chorale harmonisation~\cite{hadjeres2017deepbach}, multi-track generation using piano-rolls~\cite{dong2018musegan,huang2018music} or lead sheets~\cite{pachet2017sampling}, and modifying a given piece of music with style transfer~\cite{brunner2018midi}.

In this work, we focus on generating lead sheets from scratch. A lead sheet is a form of musical notation that represents the fundamental elements of popular songs such as chords (using chord symbols), melody and sometimes lyrics. There has been some previous research on this particular task. De Boom et al.~\cite{de2019rhythm} proposed a two-stage generation system based on long-short term memory networks (LSTMs) to produce rhythm and chord events, for which a melody sequence is generated in a second (conditioned) phase. Liu \& Yang~\cite{liu2018lead} also introduced a two-stage generation system, which generates the lead sheet first, after which a polyphonic arrangement is produced as accompaniment using Generative Adversarial Networks (GANs).

Recently, there have also been research to let the user control the output of the generation by setting some constraints. These constraints are usually referred to as ``high-level'' musical parameters and may be relatively subjective, such as style and genre. Flow Composer \cite{papadopoulos2016assisted} is an example of a conditional generative system that combines two Markov chains enriched by regular constraints, whereby the user sets the desired style of the lead sheet by selecting a corpus of existing lead sheets. \cite{pachet2017sampling} proposed an approach whereby structured lead sheets are generated, based on a mechanism (i.e.,~belief propagation) for efficiently sampling variations of existing musical sequences. In this system, the user can control certain parameters (e.g., similarity).


Extending this strategy of controlling the generation through high-level constraints imposed by the user, our system allows the user to control the emotion or valence of the generated music. Emotions and music are intrinsically connected~\cite{meyer2008emotion}, yet the qualities of the music that give rise to emotions are difficult to capture \cite{cheuk2020regression}. In order to create a training dataset, we first propose an approach for calculating the perceived emotions of the music from existing chord progressions. This allows us to label a training dataset of lead sheets with emotions, which we can then leverage to train the proposed conditional generative model. 

Our approach for calculating emotions from chords is based on~\cite{juslin2011handbook,cho2016music,zhao2019emotional} who indicate that the mode or type (e.g., major, minor, seventh, etc.) of chords corresponds directly with the valence of the music. High-level musical qualities such as emotions ``suffer'' from abstractness and subjectivity due to the fact that they require human annotations. In this work, however, instead of using human annotated labels, we filtered mood tags that correspond to different types of chords and use those to create a label for perceived emotions. The chosen tags were based on annotations by music experts~\cite{chase_2006}. In Section \ref{sec2} we leverage this knowledge and propose a novel way of annotating chords with valence. To the best of our knowledge, there is no existing work that offers a method to manually calculate the valence of a chord progression.

We use the calculated valence as a high-level conditioning feature, along with others (e.g., time signature and grouping indicators inspired by~\cite{makris2019conditional}), in our proposed generative lead sheet system based on \textit{sequence-to-sequence} architectures (LSTM~\cite{cho2014learning} and Transformer~\cite{vaswani2017attention}). Another novel aspect of our approach is a unique strategy to include the high-level user conditions as the encoder input, whereas the musical events of the lead sheet are predicted in the decoder. Thus, we approach the task of lead sheet generation much like a Neural Machine Translation (NMT) problem, except that we are translating `conditions' into `musical events'. 

The remainder of this paper is organised as follows: Section~\ref{sec2} presents our proposed strategy for calculating the valence of chord progression within a lead sheet. Next, Section~\ref{sec3} shows the details of the proposed representation for conditional lead sheet generation. Sections~\ref{sec4} \& \ref{sec5} detail the experimental setup and the evaluation of our approach, followed by a conclusion in Section~\ref{sec6}.

\section{Novel way to Measure Valence of Chords}\label{sec2}

A chord is defined as a set of two or more simultaneously played notes. A chord progression is a sequence of consecutive chords and often refers to the harmony of a song. It is considered to be a fundamental element of music that often influences the emotions a listener perceives~\cite{juslin2011handbook}. This point becomes obvious when you listen to a new arrangement of a song in which the melody remains the same but the harmony is changed. The same piece can convey entirely different emotions if the chord progression has changed~\cite{cho2016music}. \cite{lahdelma2016single} mention the importance of certain chord types inside a progression for making the music sound ``emotional''. In addition, many studies show that major chords convey positive emotion and minor chords convey negative emotion (e.g.,~\cite{bakker2015musical,pallesen2005emotion}).

We will be representing the emotion of chords in terms of valence. Valence relates to the positivity or negativity of the emotion conveyed by a song and falls on a scale from positive ($+1$) to negative ($-1$) \cite{russell1980circumplex}. For instance, anger, fear, and sadness all have low (negative) valence. On the other hand, emotions such as happy, content, and joyful  correspond to high (positive) valence. Given the fact that chord types have a direct impact on the valence of a piece~\cite{juslin2011handbook,cho2016music}, we focus here on the effect that the chord progression of a lead sheet has on perceived valence. It is worth noting that `arousal' (which refers to the energy level conveyed by the song)~\cite{russell1980circumplex} is excluded in this approach due to the absence of tempo markings in our training dataset, given that arousal is typically strongly affected by tempo~\cite{coutinho2011musical}.

To the best of our knowledge, there is only one dataset in the symbolic domain that contains  valence and arousal annotations: the VGMIDI dataset~\cite{ferreira2019learning}. This dataset contains piano arrangements of 95 video game soundtracks in MIDI, annotated with valence and arousal values in the range of $-1$ to $+1$. Although the annotations are continuous, per beat, the input files are not close to the form of a lead sheet. There are also many occasions where harmony (i.e., chord progression) cannot be identified due to the unquantified nature of the midi files. 

Thus, we propose a new method to \textbf{manually} calculate the valence of chords based on mood tags, as it is easier to find datasets with annotated mood tags, versus annotated valence and arousal. \cite{chase_2006} found a relationship between modes of chords and associated evoked emotions from professional composers and musicians (see Table \ref{table1}). This list of chords with associated emotions has already shown to be useful in related research areas such us mood classification~\cite{schuller2010determination}. We leverage these findings and consider the modes and types of chords (major, minor, 7th, etc.) in our work, because this property has arguably the greatest influence on perceived mood~\cite{lahdelma2016single}. 

\begin{figure*}[htb]
\centering
\includegraphics[width=0.8\textwidth]{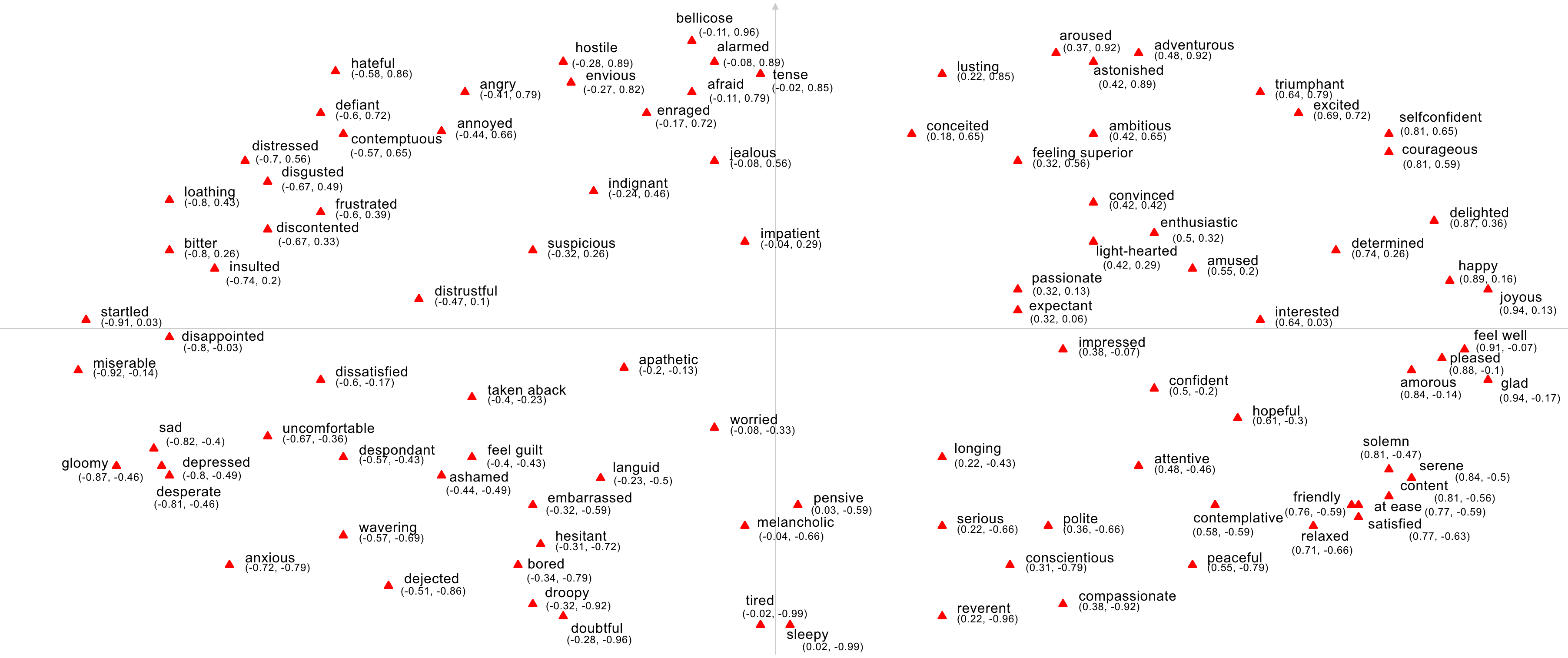}
\caption{Scherer~\cite{scherer2005emotions}'s mapping of mood tags to the two-dimensional circumplex space model proposed by Russell~\cite{russell1980circumplex}. The corresponding valence (x-axis) and arousal (y-axis) values are based on the coordinates provided by the mapping.}
\label{fig:space_model}
\end{figure*}

We then use a mapping from Paltoglou and Thelwall~\cite{paltoglou2012seeing} that matches Scherer's~\cite{scherer2005emotions} emotion tags with corresponding valence and arousal values. This allow us to retrieve a valence value for the emotion tags for each chord type from Table \ref{table1}. A representation of this mapping is shown in  Figure~\ref{fig:space_model}, where the valence and arousal values are represented on the x-axis and y-axis respectively (in the range of $-1$ to $+1$).  


\begin{table}[]
\centering
\caption{Chord types and their associated emotions, adapted from ~\cite{chase_2006}.}
\begin{tabular}{ll}
\toprule
\textbf{Chord Type (example)} & \textbf{Associated Emotions}                                                                                               \\
\midrule
Major (C)                     & \begin{tabular}[c]{@{}l@{}}Happiness, cheerfulness, \\ confidence, brightness, \\ satisfaction\end{tabular}                \\
Minor (Cm)                    & \begin{tabular}[c]{@{}l@{}}Sadness, darkness, sullenness, \\ apprehension, melancholy, \\ depression, mystery\end{tabular} \\
Dominant Seventh (C7)           & \begin{tabular}[c]{@{}l@{}}Funkiness, soulfulness, \\ moderate edginess\end{tabular}                                       \\
Major Seventh (Cmaj7)          & \begin{tabular}[c]{@{}l@{}}Romance, softness, jazziness, \\ serenity, tranquillity, \\ exhilaration\end{tabular}           \\
Minor Seventh (Cm7)             & \begin{tabular}[c]{@{}l@{}}Mellowness, moodiness, \\ jazziness\end{tabular}                                                \\
Dominant Ninth (C9)             & Openness, optimism                                                                                                         \\
Diminished (Cdim)               & \begin{tabular}[c]{@{}l@{}}Fear, shock, spookiness, \\ suspense\end{tabular}                                               \\
Suspended Fourth (Csus4)        & Delightful tension                                                                                                         \\
Seventh Minor Ninth (C7b9)      & \begin{tabular}[c]{@{}l@{}}Creepiness, ominousness, fear, \\ darkness\end{tabular}                                         \\
Added Ninth (Cadd9)             & Steeliness, austerity      \\                                \bottomrule   
\end{tabular}
\label{table1}
\end{table}

One issue we encountered when doing the mapping is that not all of the emotion tags from Table \ref{table1} were included in Scherer's model. To address this, we proceeded with two assumptions/modifications. First, the emotion tags that did not describe an ``actual'' emotion (e.g., jazziness) were removed. Next, the remaining tags that were not present in Scherer \cite{scherer2005emotions}'s model were matched as best as possible to synonyms that were present, with the help of a Music Psychologist who is a native English speaker. This resulted in a cleaned, reduced emotion tag list per chord type. 

Using this method, we can find valence values based on chords by looking at the (cleaned) emotion tag list for each chord. The final valence value is the median valence of all the descriptive cleaned emotion tags. For instance, ``Major'' has five associated emotions. ``Cheerfulness'' was matched with ``happiness'' since they are synonyms, and ``brightness'' (which was not in \cite{scherer2005emotions}'s model)  was mapped to ``delighted''. The extracted Valence values from Figure~\ref{fig:space_model} are $0.89$, $0.89$, $0.51$, $0.87$ and $0.77$ respectively, resulting in a final valence value of $0.87$. This value makes intuitive sense because a major chord is often related to a positive mood.

\section{Proposed Framework}\label{sec3}


In this section we propose a novel architecture for a ``controllable'', affective music generation system using high-level musical qualities imposed by the user. Because we consider the task of lead sheet generation to be a NMT task, we make use of popular \textit{sequence-to-sequence} (seq2seq) architectures~\cite{cho2014learning}. The user defines a sequence of musical attributes (conditions) in the $encoder$ stage, which is then ``translated'' to a complete lead sheet in the $decoder$ stage. Thus, we use two different input representations for both stages that are inspired by the event-based token representation from~\cite{oore2020time}.

\begin{figure*}[htb]
\centering
\includegraphics[width=0.75\textwidth]{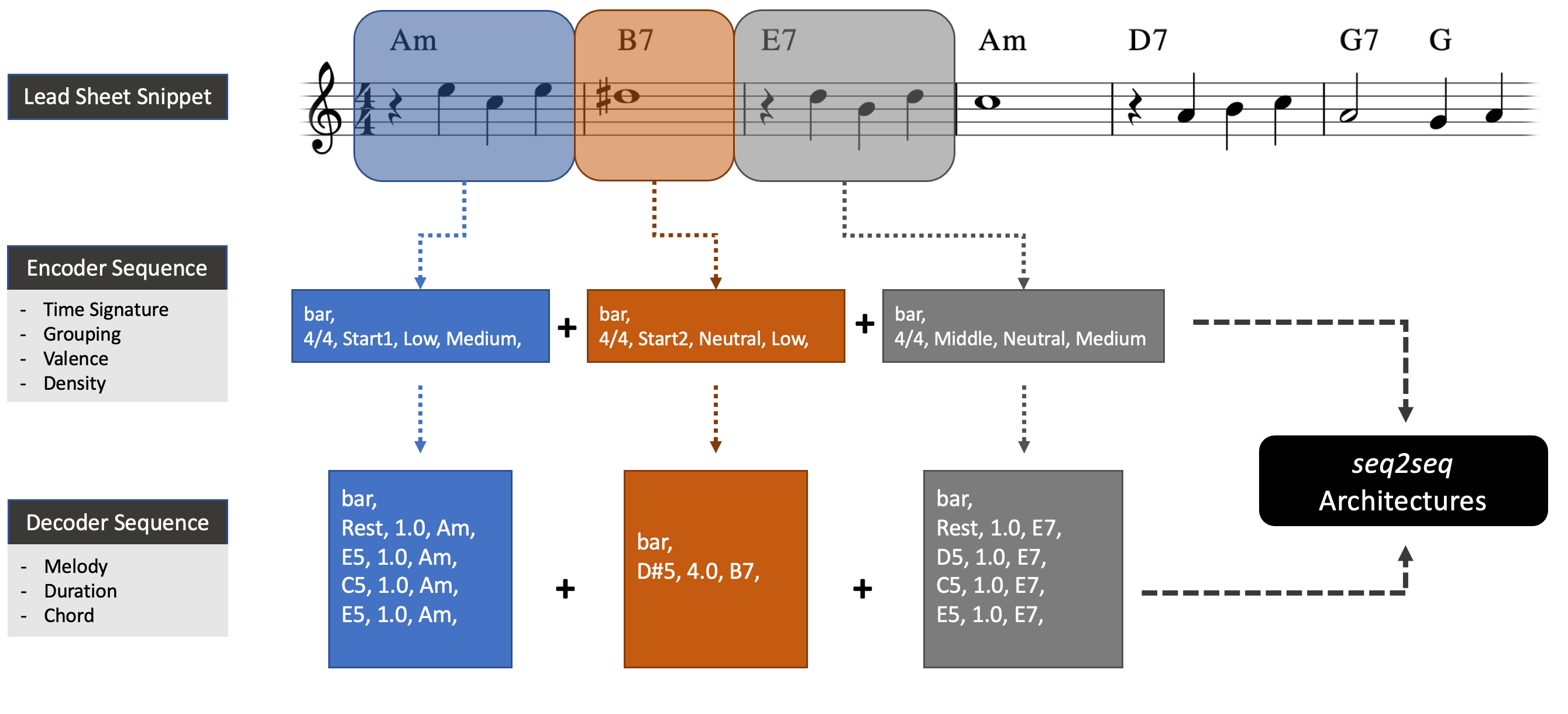}
\caption{Illustrated workflow of a lead sheet snippet transcribed to the proposed event representation for our encoder and decoder.}
\label{fig:encoding_model}
\end{figure*}

\subsection{Encoder Representation - Controllable by the user}\label{sec3.1}

In our proposed approach, the \textbf{Encoder} takes a sequence of high-level musical parameters (conditions) as input, which allows the user to guide the generated music. The system allows for a high level of control, as the user can set the desired levels of the parameters for every bar. It is possible to have either varying or constant values of these parameters during generation. The following parameters are included: 

\begin{itemize}
    \item \textbf{Chord Valence}: During training, the valence for each chord type is calculated from the training set as described in Section~\ref{sec2}. To define the overall valence within a bar, we calculate the median for each chord inside that bar. Because valence ranges between -1 to 1, discretisation is needed. We use five discrete labels (``Low'', ``Moderate Low'', ``Neutral'', ``Moderate High'' and ``High'' respectively) which are used to divide the valence range equally. Thus, chord types with close valence values will get the same descriptors.
    \item \textbf{Time Signature}: Symbols are used to describe the number of beats the ``meter'' of each bar of music~\cite{chase_2006}. Meter refers to the recurring pattern of accents that provide the pulse or beat of the music. Indicating and adjusting the time signature has a direct affect on the rhythm of the piece.
    \item \textbf{Grouping Indicators}: This feature contains extracted annotations which mark structurally coherent temporal regions of the music (i.e., different verses or refrains). We have five distinct symbols: two indicating the first two bars of a phrase, two for the last two bars and one more descriptor for the rest bars. These indicators allow the model to learn the initial and final events of a phrase, an approach that has been successfully used in the task of melodic harmonisation~\cite{tsougras2016learning} and drums generation~\cite{makris2019conditional}.
    
\end{itemize}

In addition to these high-level features, we implement a ``low-level'' feature called \textbf{event density} which is calculated as the number of events (either chord or melody) per bar. Note density will allow the user to control the number of generated events within a bar. The average number of events in the dataset used in our experimental setup (see Section~\ref{sec4.1}) is 3.68 with a variance of 2.02 per bar. We use three discrete labels: one for low, medium and high note density with ranges of $[0-2]$, $[3-5]$, and $[6+]$ events, respectively.

Given the above described conditional input, the \textbf{Encoder} sequence can be defined as:

\[
	Enc_{seq} = (bar,\ h_1,\ bar,\ h_2,\ \dots \ ,\ h_n)
\]
whereby the $bar$ event declares the start of a new bar, and $h_i$, is a vector of length 4 that represents the proposed high-level and low-level features for each bar $i$ (with $n$ being the total number of bars):
\[
	h_{\text{i}} = (TimeSig_i,\ Grouping_i,\ Valence_i ,\ Density_i)
\]

\subsection{Decoder Representation - Lead Sheet Output} \label{sec3.2}

The \textbf{Decoder} outputs the generated sequence of lead sheet events per bar. This output can be formulated as follows:

\[
	Dec_{seq} = (bar,\ l_1,\ bar,\ l_2,\ \dots \ ,\ l_n)
\]

whereby the $bar$ event declares a new bar, and $l_i$ represents a series of lead sheet events (with $n$ being the total number of bars defined in the Encoder stage) for the corresponding bar. Each lead sheet event consists of a chord symbol, melody pitch, and duration token. These tokens share the same dictionary that is built from the training dataset. The number of lead sheet events is variable for each bar, and can be controlled by the $Density_i$ feature in the Encoder. Therefore this can be formalised as:
\[
	l_{\text{i}} = (c_1,\ m_1,\ d_1 ,\ c_2,\ m_2,\ d_2,\ \dots \ ,\ c_e,\ m_e,\ d_e)
\]

where $c$, $m$ and $d$ represent a chord triad, melody pitch, and duration token respectively, with $e$ indicating the generated number of events in the bar $l_i$. Figure~\ref{fig:encoding_model} illustrates an example of a lead snippet transcribed to the proposed event representation for both Encoder and Decoder stages. This proposed representation is designed to be compatible with seq2seq architectures.

\subsection{Model architectures}\label{sec3.3}

Our proposed Encoder-Decoder representation was designed to work with state-of-the-art seq2seq architectures that have been used successfully in the NMT field. We implemented the following two architectures:

\begin{enumerate}
    \item \textbf{LSTM-based encoder-decoder}: Inspired by \cite{de2019rhythm}, we implemented a 3-layer Bidirectional LSTM (BiLSTM)~\cite{graves2005framewise} Encoder, consisting of 512 hidden units and a 3-layer Decoder of 1,024 hidden units with 30\% dropout between consecutive layers. The BiLSTM states provided by the Encoder allow the lead sheet generator (i.e., Decoder) to look back as well as ahead at the sequence of music parameters defined by the user. It is worth noting that adding local or global attention mechanisms~\cite{luong2015effective} to this network did not provide a better performance during training or generation. This may be due to the nature of the input representation of the encoder which is a sequence that describes a specific bar, whereas the decoder output is a list of multiple events within bars. Therefore, the size of the decoder is much larger than the encoder. 
    \item \textbf{Transformer}: We adapted the vanilla version of the transformer architecture from~\cite{vaswani2017attention} in our proposed system. A total of 4 self-attention layers and 8 multi-head attention modules were used. The number of hidden units of the Feed-Forward layers was set to 1,536, with 20\% dropout between consecutive layers.
\end{enumerate}

\subsection{Training and Generation}\label{sec3.4}

Due to the nature of our token-based encoding representation, we tackle the training and generation procedures with ``teacher-forcing'', a technique often used for NMT or text-generation tasks. Thus, it is worth noting that both seq2seq architectures have single outputs in the decoder stage, and hence generate a single token in every iteration (inspired by the encoding representations from \cite{oore2020time,huang2018music}) and not a triplet lead sheet event as described in the previous section. Therefore, in the decoder stage we have a single dictionary that includes all the tokens for chords, melody pitches, and durations. 

We use the Adam optimizer with a learning rate of $0.001$, and categorical cross-entropy as the loss function, for both models. The batch size was set to 32 and the model was implemented using Tensorflow 2.x~\cite{abadi2016tensorflow}. In order to control the diversity of the generation, we use a temperature $\tau$ (which was varied uniformly randomly between 0.8 and 1.2) to sample from the output distribution. Finally, in the generation stage, the user can either define the parameters manually for every bar, or randomly generate control parameter templates based on the statistics of the test set. The code and pre-processed dataset are available on GitHub\footnote{\url{https://github.com/melkor169/LeadSheetGen_Valence}}.

\section{Experimental Setup}\label{sec4}
The goal of our proposed framework is to generate novel lead sheets that allow the user to control features such as valence at the bar-level. Our experiment focuses on the valence of the generated chord progressions, an approach made viable based on our unique strategy for chord valence calculation (Section~\ref{sec2}) that allows us to create a lead sheet dataset with labelled valence values. Therefore, we aim to evaluate:

\begin{enumerate}
    \item Does the valence input by the user affect the perceived valence of the generated music as intended? 
    \item How effective is our proposed encoder/decoder representation and architecture to generate high quality, `real' sounding music?
\end{enumerate}

We address these research questions by both calculating an extensive set of evaluation metrics as well as a listening test described in Section~\ref{sec:4.3}.

\subsection{Data Collection and Pre-Processing}\label{sec4.1}
For our experiments, we use the Wikifonia dataset\footnote{Wikifonia archive is no more directly accessible from the web. You may contact the authors if you are interested in obtaining the original dataset.}. This dataset contains 6,675 lead sheets in MusicXML format, including diverse genres, from folk to popular music. After filtering out corrupted files that do not contain chord symbols or melody notation, we proceeded with the following pre-processing steps:

\begin{itemize}
    \item All of the songs were transposed to the key of C major or A minor. Songs that contain key changes were split into different independent instances. In addition, there was a limit of up to 32 bars length for every song.
    \item Inspired by~\cite{de2019rhythm}, we eliminated polyphonic melody parts and ignored ties between notes from different bars. Moreover, we unfolded repetitions since lead sheets can contain repeated phrases. Therefore if a repeat barline symbol occurs, we duplicate that particular phrase.
    \item We set restrictions on the available modes and chord types. Table~\ref{table2} shows the permitted chord types to allow for our valence calculation method. Lead sheets including other chord types were removed. In addition, we removed inversions in the chord symbols.
    \item Multiple Time Signatures were detected in the original dataset, however, we only considered lead sheets with the 5 most common ones: $4/4$, $3/4$, $2/2$, $2/4$ and $6/8$.
    \item We only included lead sheets with the most frequent durations in the dataset, including triplets (of quarter notes or eighth notes). Finally, for melodic pitch, the permitted range was set from $G3$ (55) to $C6$ (84).
    
\end{itemize}

The resulting processed dataset contains 4,776 lead sheets with a variable length of 4 up to 32 bars. This was divided into training / validation / test sets with a ratio of 8:1:1. Both the Chord and Melody list of tokens include the ``Rest'' symbol.

\begin{table}[]
\centering

\caption{Occurrences of the different chord modes and their associated valence values in our dataset.}

\begin{tabular}{@{}lrr@{}}
\toprule
\textbf{Chord Type} & \textbf{Valence}                                     & \textbf{Occurrences}                                   \\
\midrule
Major               & \cellcolor[HTML]{FFFFFF}{\color[HTML]{036400} 0.87}  & \cellcolor[HTML]{FFFFFF}{\color[HTML]{000000} 333,232} \\
Minor               & \cellcolor[HTML]{FFFFFF}{\color[HTML]{9A0000} -0.81} & \cellcolor[HTML]{FFFFFF}{\color[HTML]{000000} 89,741}  \\
Dominant Seventh    & \cellcolor[HTML]{FFFFFF}{\color[HTML]{6665CD} -0.02} & \cellcolor[HTML]{FFFFFF}{\color[HTML]{000000} 173,586} \\
Major Seventh       & \cellcolor[HTML]{FFFFFF}{\color[HTML]{036400} 0.83}  & \cellcolor[HTML]{FFFFFF}{\color[HTML]{000000} 19,617}  \\
Minor Seventh       & \cellcolor[HTML]{FFFFFF}{\color[HTML]{FD6864} -0.46} & \cellcolor[HTML]{FFFFFF}{\color[HTML]{000000} 55,536}  \\
Dominant Ninth      & \cellcolor[HTML]{FFFFFF}{\color[HTML]{32CB00} 0.51}  & \cellcolor[HTML]{FFFFFF}{\color[HTML]{000000} 12,944}  \\
Minor Ninth         & \cellcolor[HTML]{FFFFFF}{\color[HTML]{6665CD} -0.15} & \cellcolor[HTML]{FFFFFF}{\color[HTML]{000000} 9,557}   \\
Diminished          & \cellcolor[HTML]{FFFFFF}{\color[HTML]{FD6864} -0.43} & \cellcolor[HTML]{FFFFFF}{\color[HTML]{000000} 9,001}  \\
\bottomrule   

\end{tabular}
\label{table2}
\end{table}

\begin{table*}[ht!]
\centering
\caption{Results of Quantitative Evaluation in terms of the proposed metrics (mean $\pm$ standard deviation).}
\begin{tabular}{c}
  \begin{tabular*}{\linewidth}{!{\extracolsep\fill}@{}
    >{\columncolor[HTML]{FFFFFF}}l 
    >{\columncolor[HTML]{FFFFFF}}c 
    >{\columncolor[HTML]{FFFFFF}}c 
    >{\columncolor[HTML]{FFFFFF}}c 
    >{\columncolor[HTML]{FFFFFF}}c 
    >{\columncolor[HTML]{FFFFFF}}c @{}}
    \toprule
    {\color[HTML]{000000} } &
    
      \multicolumn{2}{c}{\cellcolor[HTML]{FFFFFF}{\color[HTML]{000000} \textbf{Used Pitch Classes}}} &
     
      \multicolumn{2}{c}{\cellcolor[HTML]{FFFFFF}{\color[HTML]{000000} \textbf{Rest Events (\%)}}} &
      {\color[HTML]{000000} \textbf{Tonal Distance}} \\ 
    {\color[HTML]{000000} } &
      {\color[HTML]{000000} Melody} &
      {\color[HTML]{000000} Chords} &
      {\color[HTML]{000000} Melody} &
      {\color[HTML]{000000} Chords} &
      {\color[HTML]{000000} Melody - Chords} \\
    \midrule
    {\color[HTML]{000000} Training Dataset} &
      {\color[HTML]{000000} 2.5896 $\pm$ 1.1283} &
      {\color[HTML]{000000} 4.8602 $\pm$ 1.6168} &
      {\color[HTML]{000000} 0.0755 $\pm$ 0.1871} &
      {\color[HTML]{000000} 0.0132 $\pm$ 0.0574} &
      {\color[HTML]{000000} 1.4634} \\ \midrule
    {\color[HTML]{000000} Proposed LSTM-based} &
      {\color[HTML]{000000} \textbf{2.6503 $\pm$ 1.1166}} &
      {\color[HTML]{000000} 4.4447 $\pm$ 1.6251} &
      {\color[HTML]{000000} \textbf{0.0806 $\pm$ 0.1993}} &
      {\color[HTML]{000000} \textbf{0.0344 $\pm$ 0.1006}} &
      {\color[HTML]{000000} 1.6432} \\
    {\color[HTML]{000000} Proposed Transformer} &
      {\color[HTML]{000000} 2.3688 $\pm$ 1.0856} &
      {\color[HTML]{000000} 4.3101 $\pm$ 1.6335} &
      {\color[HTML]{000000} 0.0886 $\pm$ 0.2147} &
      {\color[HTML]{000000} 0.0424 $\pm$ 0.0884} &
      {\color[HTML]{000000} \textbf{1.4918}} \\
    {\color[HTML]{000000} LSTM two stages~\cite{de2019rhythm}} &
      {\color[HTML]{000000} 2.2660 $\pm$ 1.1228} &
      {\color[HTML]{000000} 4.4483 $\pm$ 1.7735} &
      {\color[HTML]{000000} 0.1079 $\pm$ 0.2499} &
      {\color[HTML]{000000} 0.0591 $\pm$ 0.1223} &
      {\color[HTML]{000000} 1.5043} \\
    {\color[HTML]{000000} MuseGAN two tracks~\cite{dong2018musegan}} &
      {\color[HTML]{000000} 2.8575 $\pm$ 1.1643} &
      {\color[HTML]{000000} \textbf{4.6541 $\pm$ 1.5931}} &
      {\color[HTML]{000000} 0.1695 $\pm$ 0.1895} &
      {\color[HTML]{000000} 0.0437 $\pm$ 0.0807} &
      {\color[HTML]{000000} 1.6543} \\ \bottomrule
    \end{tabular*} \\
    \\
    \resizebox{\linewidth}{!}{
    \begin{tabular}{@{}
    >{\columncolor[HTML]{FFFFFF}}l 
    >{\columncolor[HTML]{FFFFFF}}c 
    >{\columncolor[HTML]{FFFFFF}}c 
    >{\columncolor[HTML]{FFFFFF}}c 
    >{\columncolor[HTML]{FFFFFF}}c 
    >{\columncolor[HTML]{FFFFFF}}c 
    >{\columncolor[HTML]{FFFFFF}}c @{}}
    \toprule
    {\color[HTML]{000000} } &
      \multicolumn{1}{l}{\cellcolor[HTML]{FFFFFF}{\color[HTML]{000000} }} &
      {\color[HTML]{000000} \textbf{Pattern Metrics}} &
      \multicolumn{1}{l}{\cellcolor[HTML]{FFFFFF}{\color[HTML]{000000} }} &
      \multicolumn{1}{l}{\cellcolor[HTML]{FFFFFF}{\color[HTML]{000000} }} &
      \textbf{Tension Metrics} &
      \multicolumn{1}{l}{\cellcolor[HTML]{FFFFFF}} \\
    {\color[HTML]{000000} } &
      {\color[HTML]{000000} Compression Ratio} &
      {\color[HTML]{000000} Long Patterns (avg)} &
      {\color[HTML]{000000} Short Patterns (avg)} &
      {\color[HTML]{000000} Cloud Movement} &
      Cloud Diameter &
      Distance to the Key \\
      \midrule
    {\color[HTML]{000000} Training Dataset} &
      {\color[HTML]{000000} 1.7384 $\pm$ 0.1784} &
      {\color[HTML]{000000} 1.8039 $\pm$ 3.8745} &
      {\color[HTML]{000000} 15.3772 $\pm$ 5.9791} &
      {\color[HTML]{000000} 0.3197 $\pm$ 0.0987} &
      2.4351 $\pm$ 0.3584 &
      0.5639 $\pm$ 0.1083 \\ \midrule
    {\color[HTML]{000000} Proposed LSTM-based} &
      {\color[HTML]{000000} 1.6599 $\pm$ 0.1113} &
      {\color[HTML]{000000} 0.8823 $\pm$ 1.8601} &
      {\color[HTML]{000000} 17.5720 $\pm$ 6.0227} &
      {\color[HTML]{000000} 0.3012 $\pm$ 0.0839} &
      2.2780 $\pm$ 0.3426 &
      0.5592 $\pm$ 0.1102 \\
    Proposed Transformer &
      \textbf{1.7654 $\pm$ 0.2185} &
      \textbf{2.1533 $\pm$ 3.9989} &
      \textbf{14.4458 $\pm$ 5.9062} &
      0.2742 $\pm$ 0.0994 &
      2.2545 $\pm$ 0.3287 &
      \textbf{0.5667 $\pm$ 0.1136} \\
    LSTM two stages~\cite{de2019rhythm} &
      1.6715 $\pm$ 0.1267 &
      0.8190 $\pm$ 1.8318 &
      16.9420 $\pm$ 5.9048 &
      \textbf{0.3168 $\pm$ 0.1006} &
      2.2266 $\pm$ 0.4118 &
      0.6101 $\pm$ 0.1098 \\
    MuseGAN two tracks~\cite{dong2018musegan} &
      1.5355 $\pm$ 0.0664 &
      0.2245 $\pm$ 1.0132 &
      23.8170 $\pm$ 6.5830 &
      0.2698 $\pm$ 0.2065 &
      \textbf{2.4879 $\pm$ 0.6753} &
      0.6047 $\pm$ 0.1774\\
      \bottomrule
    \end{tabular}
    }
\end{tabular}

\label{table3_set1}
\end{table*}

\subsection{Evaluation metrics}\label{sec:4.2}
We conduct both a computational experiment and a user study to evaluate the quality of music generated by our proposed method. To evaluate how well our approach can create emotionally distinctive music based on the valence of the chord progression, we examine whether the participants in the listening study are able to identify the overall perceived emotions of the generated lead sheets.

\subsubsection{Analytical measures}
There is no standard way to quantitatively measure whether a lead sheet generation model has been trained well \cite{agres2016evaluation}. However, we adopt the following measures that were recently proposed in~\cite{dong2018musegan} and have been used to evaluate lead sheet generation from scratch~\cite{liu2018lead}.

\begin{itemize}
    \item \textbf{Used Pitch Classes}: Average number of used pitch classes per bar for both melody and chord tracks.
    \item \textbf{Rest Events}: This is a modification of the proposed ``Empty Bars'' metric, as we do not encounter empty bars in our training dataset. Thus, this metric indicates the average ratio of rest events per bar for both melody and chord tracks.
    \item \textbf{Tonal Distance}: Measures the harmonicity between two given tracks~\cite{harte2006detecting}. Large values of Tonal Distance implies weaker inter-track harmonic relations between the Melody and the Chord track.
\end{itemize}

In addition, we propose two sets of calculated measures that can be used in the objective evaluation of generated lead sheets (see ~\cite{chuan2018modeling, guo2021hrnn}). First, by measuring the ``compression ratio'' of generated content we can measure the number of repeated patterns, which is related to ``long-term structure''. This can be calculated using the \textbf{Omnisia}\footnote{\url{https://github.com/chromamorph/omnisia-recursia-rrt-mml-2019}} software which uses the COSIATEC compression greedy algorithm~\cite{meredith2013cosiatec} and computes the following metrics:

\begin{itemize}
    \item \textbf{Compression Ratio}: A measure of detecting repeated patterns such as themes and motives in the generated musical content.
    \item \textbf{Average Long Patterns}: Measures the average number of the longest detected patterns (i.e.,~in terms of note events) within a lead sheet.
    \item \textbf{Average Short Patterns}: Indicates the average number of the shortest detected patterns.
\end{itemize}

Finally, we calculate tension measures proposed by~\cite{herremans2016tension, herremans2017morpheus} to quantify the tension profile of a musical song. Musical tension forms an essential part of the experience of listening to music -- increased tension levels can be subjectively described as ``a feeling of rising intensity'', while decreased tension is a ``feeling of relaxation''~\cite{farbood2012parametric}. We calculate the following measures which are based on the spiral array proposed by~\cite{chew2002spiral}:

\begin{itemize}
    \item \textbf{Cloud Diameter}: Indicates the level of dissonance within a sliding window frame (i.e.,~``cloud'').
    \item \textbf{Cloud Momentum}: Measures the distance (tonality movement) between different clouds
    \item \textbf{Tensile Strain}: Calculates the tonal distance between a cloud of notes and the key of the piece.
\end{itemize}

\subsubsection{Listening test setup}\label{sec:4.3}

We conducted an online listening test in which participants rated 15 short samples of lead sheets ranging from 20 to 40 seconds in duration. Each sample was presented in the form of a video clip which captures the playback of a  lead sheet, so that the user could also see the chord symbols. The distribution of the samples was as follows:
(i) 5 samples selected randomly from the test set, (ii) 5 samples generated with the Transformer architecture, and (iii) another 5 samples generated using the LSTM-based architecture. 

First, we wanted to subjectively measure which proposed model sounds more ``pleasant'' and coherent. Thus, each participant was asked to rate each sample on a 5-point Likert scale, ranging from 1 (very low) to 5 (very high), using four criteria that were adapted from ~\cite{dong2018musegan,liu2018lead}:
\begin{enumerate}
    \item Rhythm: Whether the Rhythm events are pleasant.
    \item Melody: How novel the generated Melody is.
    \item Harmony: If the Chord Progression sounds coherent.
    \item Naturalness: Whether the ``humanised'' element is perceived.
\end{enumerate}

Second, in order to evaluate whether our proposed approach can really steer the valence of the generated chord progression, we asked the participants to rate their overall perceived valence of the chord progressions, using the five discrete labels from Section~\ref{sec3.1} which we refer to as valence descriptors in this experiment. For the test samples we calculated the average valence using our novel method presented in Section~\ref{sec2}. These valence values were then used as input conditions to generate music pieces. These new pieces were then rated by listeners in terms of valence. This allows us to evaluate whether the desired (input) valence is the same as the valence perceived by actual human listeners.

\section{Experimental Results}\label{sec5}

\subsection{Quantitative Evaluation}\label{sec:5.1} 
We compare the evaluation scores of our proposed method with two related state-of-the-art approaches. Specifically:
\begin{itemize}
    \item \textbf{LSTM - Two stages}: We re-created the two-stage LSTM model from~\cite{de2019rhythm} with the same configuration and hyper-parameters. In the first stage, Rhythm and Chord events are generated together using two stacked LSTM layers. Next, the previous output is fed to the BiLSTM layers to get the states and generate the Melody with two stacked LSTM layers again.
    \item \textbf{MuseGAN - Two tracks}: We adapted the model proposed by~\cite{liu2018lead} which generates Lead Sheets from scratch as a first stage using the MuseGAN~\cite{dong2018musegan} architecture, a multi-track Sequential Generative Adversarial Network. We reduced the generated tracks to Melody and Chords only, and converted our training data to piano-roll format (an alternative symbolic representation) to be compatible with the network input. 
\end{itemize}

Moreover since MuseGAN cannot generate sequences of variable length, we set a fixed length of 8 bars in a 4/4 Time Signature. Therefore, the training data was split into phrases in order to train all the models. We generated a total of 2,000 sequences. For our two proposed models (i.e., Transformer and LSTM-based), we used random sequences to act as Encoder inputs that were generated from  normal distributions of the corresponding musical parameters which were acquired from the training dataset. Finally, the hyper-parameter temperature $\tau$ was fixed at 1.0 for all models.

Table~\ref{table3_set1} shows the results of all the proposed metrics for all the models. Values close to those extracted from the training data indicate that the generated fragments may have more chance to be musically valid as they match the properties of existing music. Regarding the first set of metrics proposed by~\cite{dong2018musegan}, we can observe that the proposed LSTM-based seq2seq model achieves the best results in almost all categories except for the Tonal Distance and the average Used Pitch Classes for Chords. However, the small mean differences and high standard deviations in all models (even in the training set) suggest that strong conclusions cannot be made. One reasonable explanation for the high standard deviations may be the fact that Used Pitch Classes and Rest Events metrics are computed per bar in which the density of events can have large fluctuations.

Surprisingly, there is a huge difference in the pattern metrics. Based on these results, it seems that the Transformer is able to better generate sequences with long-term structure. This highlights the effectiveness of our proposed representation, which seems to work better in the Transformer architecture than the LSTM-based model. In addition, MuseGAN  fails to produce repeated patterns, which can be explained by the small number of training instances and the nature of the piano-roll representation. Finally, regarding the tension measures, the two baseline models seem to have slightly better results but, once again, the difference is quite small, especially if we take into account the standard deviation.


\begin{table}[]
\caption{Listening experiment ratings (mean $\pm$ 95\% Confidence Interval) for pieces generated by the two proposed architectures as well as existing (human) compositions.}
\begin{tabular}{@{}
>{\columncolor[HTML]{FFFFFF}}c 
>{\columncolor[HTML]{FFFFFF}}c 
>{\columncolor[HTML]{FFFFFF}}c 
>{\columncolor[HTML]{FFFFFF}}c 
>{\columncolor[HTML]{FFFFFF}}c @{}}
\toprule
\multicolumn{1}{l}{\cellcolor[HTML]{FFFFFF}} &
  \textbf{Rhythm} &
  \textbf{Melody} &
  \textbf{Chords} &
  \textbf{Naturalness} \\ \midrule
\begin{tabular}[c]{@{}c@{}}Human\\Composer\end{tabular} &
  \textbf{3.62 $\pm$ 0.16} &
  3.50 $\pm$ 0.17 &
  3.56 $\pm$ 0.16 &
  \textbf{3.64 $\pm$ 0.16} \\
LSTM-based & 3.47 $\pm$ 0.17 & 3.43 $\pm$ 0.17          & 3.51 $\pm$ 0.15          & 3.28 $\pm$ 0.15 \\
Transformer  & 3.53 $\pm$ 0.14 & \textbf{3.68 $\pm$ 0.16} & \textbf{3.76 $\pm$ 0.14} & 3.41 $\pm$ 0.17 \\ \bottomrule
\end{tabular}
\label{table4}
\end{table}

\subsection{Subjective Listening Test}\label{sec:5.2}
A total of 42 subjects participated in our listening test. All of the participants indicated that they have a strong musical background and a profession related to the music industry (e.g., composers, producers, performers, and music information retrieval (MIR) researchers). We targeted these groups because we believe that our specialised experimental questions may have been confusing for users without sufficient musical knowledge. In an overall of 630 votes, Table~\ref{table4} reveals that both proposed models' ratings on the different aspects of musicality of the piece are very close to those given to the real compositions. In addition, the Transformer seems to generate ``more coherent'' and ``more pleasant'' compositions than the LSTM-based model, and may outperform the real compositions when it comes to melody and chords.

The effectiveness of our proposed method to generate music with a particular valence is shown in Table~\ref{table5}. This table shows the valence descriptors that were given as input conditions for generation (or, for the human pieces, those calculated manually), as well as the valence descriptors from averaged participant ratings in the experiment. From the 15 musical fragments in the experiment, 11 were matched correctly with the corresponding input valence descriptor. The 4 mismatches may be due to the level of subjectivity involved in labelling valence. Interestingly, we find that almost all mismatches are those on the real human compositions.

\begin{table}[]
\centering
\caption{Subjective Evaluation of the ability of the proposed method to successfully generate a Chord Progression according to the Calculated Valence Descriptor. The colour of the User Valence Descriptor indicates the success (green) or failure (red) of the model, i.e. a match between the input condition and the resulting users' average rating.} 
\begin{tabular}{@{}
>{\columncolor[HTML]{FFFFFF}}c 
>{\columncolor[HTML]{FFFFFF}}c 
>{\columncolor[HTML]{32CB00}}c 
>{\columncolor[HTML]{FFFFFF}}c @{}}
\toprule
\textbf{Track} &
  \textbf{\begin{tabular}[c]{@{}c@{}c@{}}Calculated Valence\\ Descriptor \\(input cond.)\end{tabular}} &
  \cellcolor[HTML]{FFFFFF}\textbf{\begin{tabular}[c]{@{}c@{}c@{}}User Valence \\ Descriptor\\ (of output)\end{tabular}} &
  \textbf{Model} \\ \midrule
\textbf{1}  & Neutral       & Neutral                               & Human Composer \\
\textbf{2}  & Moderate High & Moderate High                         & Transformer    \\
\textbf{3}  & Neutral       & Neutral                               & LSTM - based   \\
\textbf{4}  & Moderate High & Moderate High                         & Human Composer \\
\textbf{5}  & Neutral       & \cellcolor[HTML]{F56B00}Moderate High & Transformer    \\
\textbf{6}  & Moderate Low  & Moderate Low                          & LSTM - based   \\
\textbf{7}  & High          & \cellcolor[HTML]{F56B00}Moderate High & Human Composer \\
\textbf{8}  & Moderate High & Moderate High                         & Transformer    \\
\textbf{9}  & Moderate Low  & Moderate Low                          & LSTM - based   \\
\textbf{10} & Neutral       & \cellcolor[HTML]{F56B00}Moderate Low  & Human Composer \\
\textbf{11} & Moderate Low  & Moderate Low                          & Transformer    \\
\textbf{12} & High          & \cellcolor[HTML]{F56B00}Neutral       & Human Composer \\
\textbf{13} & Moderate High & Moderate High                         & LSTM - based   \\
\textbf{14} & Moderate Low  & Moderate Low                          & Transformer    \\
\textbf{15} & Moderate High & Moderate High                         & LSTM - based   \\ \bottomrule
\end{tabular}
\label{table5}
\end{table}

\section{Conclusions}\label{sec6}
This paper introduces a novel strategy for conditional lead sheet generation that allows the user to steer the music generation using high-level musical qualities. We present a novel approach for calculating the valence of a chord progression using pre-defined mood tags proposed by music experts. Then, by tackling the task of lead sheet generation as a Neural Machine Translation problem, we propose a new approach to represent musical conditions such as valence as input to the $encoder$ stage of popular \textit{sequence-to-sequence} architectures. These conditions are then translated into musical lead sheet events in the $decoder$ stage. An analytical experiment and listening test  show that the proposed strategy is able to produce lead sheets in a controllable manner with similar musical attributes to the training dataset, that contain long-term structure, and which sound coherent and pleasant. In addition, the results of the listening test indicate the effectiveness of the proposed strategy to calculate and control the valence of a generated chord progression. In future research, we might examine the effect of using a noise-robust loss function (e.g.~\cite{zhang2018generalized}) on the model performance, along with adding more high-and low-level musical conditions such us arousal and just-intonation.
\bibliographystyle{IEEEtran}
\bibliography{biblio}

\end{document}